# Side-Channel VoIP Profiling Attack against Customer Service Automated Phone System

Roy Laurens[1], Edo Christianto[2], Bruce Caulkins[3], Cliff C. Zou[4]

[1,4]Department of Computer Science, University of Central Florida, Orlando, FL, USA
[2]Department of Electrical and Computer Engineering, University of Central Florida, Orlando, FL, USA
[3]School of Modelling, Simulation and Training, University of Central Florida, Orlando, FL, USA
[1,2]Department of Computer Engineering, Dinamika University, Surabaya, Indonesia
[1]rlaurens@knights.ucf.edu, [2]edo@knights.ucf.edu, [3]bruce.caulkins@ucf.edu, [4]czou@cs.ucf.edu

*Abstract* — **In many VoIP systems, Voice Activity Detection (VAD) is often used on VoIP traffic to suppress packets of silence in order to reduce the bandwidth consumption of phone calls. Unfortunately, although VoIP traffic is fully encrypted and secured, traffic analysis of this suppression can reveal identifying information about calls made to customer service automated phone systems. Because different customer service phone systems have distinct, but fixed (pre-recorded) automated voice messages sent to customers, VAD silence suppression used in VoIP will enable an eavesdropper to profile and identify these automated voice messages. In this paper, we will use a popular enterprise VoIP system (Cisco CallManager), running the default Session Initiation Protocol (SIP) protocol, to demonstrate that an attacker can reliably use the silence suppression to profile calls to such VoIP systems. Our real-world experiments demonstrate that this side-channel profiling attack can be used to accurately identify not only what customer service phone number a customer calls, but also what following options are subsequently chosen by the caller in the phone conversation.**

*Keywords—VoIP; side-channel attack; automated phone system*

## I. Introduction

Voice over Internet Protocol (VoIP) offers significant advantage compared to traditional circuit-switched voice networks. With the circuit-based call, a dedicated 64-Kbps fixed bandwidth link is required regardless of how much of the call is speech or how much is silence. A VoIP call, on the other hand, packetize all the conversation. Therefore, it can suppress the packets of silence, called Voice Activity Detection (VAD), where up to 35 percent bandwidth savings can be obtained [1]. These bandwidth savings can then be used for other network application, which makes VoIP more efficient compared to the circuit-based solution.

However, this silence suppression has unintended consequences with significant privacy implication. A cycle of voice traffic stream and silence creates a distinct pattern that can be identified and catalogued. This pattern exists even if the data stream itself is encrypted, and the actual IP-phone end point is unknown. VoIP traffic uses a codec to encode/decode the voice, and each codec uses a specific packet size and interval, and the presence of the cycle of voice traffic stream and silence cannot be obfuscated by encryption.

Normally, human conversations have enough variability in their speech pattern, speed, etc., that makes silence analysis impractical as an attack vector because even the same person will not say the same words in exactly the same way every time [2]. However, if the call is made to an automated customer service phone system (usually a toll-free 1-800 numbers), then it will be answered by an Interactive Voice Response (IVR) recording, which has a constant pattern of speech and silence due to its fixed and pre-recorded voice messages. Therefore, by profiling and cataloging the calls to such a customer service automated phone system, we can reliably identify whether subsequent calls are made to the number that we have profiled. Furthermore, we can even identify the subsequent options that the caller selects during the call using the same method.

In this paper, we set up a VoIP testbed and use it to collect and fingerprint the VoIP traffic of various popular customer service automated phone systems, such as Walmart, airlines, banks, insurance companies, etc. We demonstrate that these customer service automated phone systems have clearly distinguished voice messages that make it very easy to profile and thus are vulnerable to the presented side-channel profiling attack. Our real-world experiments show that an eavesdropper can accurately identify not only what customer service phone number a customer calls, but also what following options are subsequently chosen by the caller in the phone conversation.

## II. Related Work

### A. VoIP Attack

Research on VoIP attacks are mostly focused on the call setup stage of the protocol. The technical details of the attacks are different, but the methods are more or less the same. Ghafarian et al. [3] showed that VoIP protocols can be attacked with Denial of Service (DoS). They set up a VoIP environment and launched DoS flood attack to the SIP server. As VoIP is employed on the top of IP infrastructure, security of other protocols such as DNS, DHCP, TLS/SSL, and routing protocols, among others, must also be implemented properly. Failure to do so affects VoIP critically. By targeting vulnerable protocols and signaling at the outset of call setup, rerouting calls or interception is also possible. Wang et al. [4] investigated a man-in-the-middle (MITM) attack, specifically VoIP call diversion to a bogus IVR or representative. Unlike our paper, all the attacks described in these papers assumes that the VoIP messages are unencrypted, whereas our attack is feasible even for encrypted calls.

### B. Traffic analysis

Analysis of network traffic can reveal private information, for example, Alyami et al. [5] studied a privacy attack in which the profiling is performed using IoT devices' network traffic monitored from out-of-network. With regards to voice conversation, the analysis can be divided into active (i.e., probing) attack and passive (i.e., eavesdropping).

As an example of active attack, Shintre et al. [6] send continuous probes to targets and analyze the response traffic to reveal which target is calling one another. However, this will only work on private network.

For passive attack, there is plenty of work that looks at both the packet length, such as Wright et al. [7] and Dupasquier et al. [8], whereas Lella [9] looks at the silence suppression, which is similar to our approach. But all of them focused on identifying words or phrases that were specifically created and spoken just for the test. In contrast, our paper targets real customer service phone systems that has practical impact. We are also testing against hardware-based VoIP that is widely used in the real world, instead of app-based VoIP (i.e., WhatsApp, etc.) that cannot make a call to ordinary phone number.

### III. THREAT MODEL AND TESTBED OVERVIEW

In this section, we will give a short introduction about VoIP, followed by an explanation of the characteristic of the attacker and the testbed that we built.

#### A. VoIP Primer

VoIP, also known as IP Telephony, is the transmission of voice signals using Internet Protocol (IP) over the data network, such as the Internet. IP Phones that runs VoIP service will encode the incoming voice signal into data stream for transmission, and it will also decode incoming voice data stream back into its original audio signal. There are various supported encoder/decoder (codec) for VoIP [1], some of which are shown in table 1. All these codecs have a constant interval between voice payload, which make them susceptible to our attack.

TABLE I. VOIP CODEC

| Codec | Voice Payload (Bytes) | Voice Payload Interval | Packet Per Second (PPS) |
|---|---|---|---|
| G.711 | 160 Bytes | 20 ms | 50 |
| G.729 | 20 Bytes | 20 ms | 50 |
| G.723.1 | 24 Bytes | 30 ms | 33.3 |
| G.723.1 | 20 Bytes | 30 ms | 33.3 |
| G.722 | 160 Bytes | 20 ms | 50 |

In addition to compatible codec, VoIP also need a signaling protocol so the IP phones can dial one another. The current industry standard for this is Session Initiation Protocol (SIP), which is an RFC standard from Internet Engineering Task Force (IETF) to establish sessions in an IP network [10]. SIP operates in a client-server model, with a SIP server facilitating signalling between VoIP phones that wants to start a communication session. The SIP server also often serves as a VoIP gateway to Public Switched Telephone Network (PSTN) so VoIP phones can call and communicate with traditional phones.

Since VoIP runs on top of existing data network, the VoIP phones and SIP server/gateway does not have to share the same physical location. In a corporate environment, a SIP server can be located in the headquarter, while the IP phones in the branch locations connect to it using existing Virtual Private Network (VPN) data link, for example. There are also many companies offering cloud phone system, where they offer to host the SIP server functionality and the customer can connect his/her IP phone to this server via the Internet. Although these voice data streams might be encrypted, the fixed size and constant interval of the stream, combined with the silence suppression (VAD) makes this transmission susceptible to the attack outlined below.

#### B. Threat Model

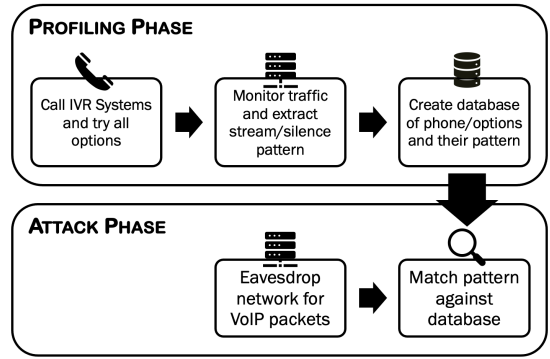

Fig. 1. VoIP attack threat model

Our attack exploits the patterns of traffic streams and silence, so the attacker will first need to create a database that catalogues the distinct stream and silence characteristic of various customer service toll-free numbers (Fig 1). This can be done by making the call by the attacker himself and extracting the packet stream features of that call. Of course such a database cannot completely cover all existing customer service phone numbers, so we will have a category of 'unclassified' entry to denote unknown numbers. However, even the unclassified entry can reveal private information because we can identify if calls are made to the same IVR system, and whether the same IVR options are chosen, even if we don't know what phone numbers are being called.

Once the database is built, the attacker can infer sensitive information about a VoIP phone call to an automated customer service phone system, by eavesdropping on the traffic stream of such a call. The observed packet stream and silence pattern will be matched against known pattern to identify the phone number and the options chosen by the caller. The sniffing can happen at any path that the VoIP data stream traverse, so we can expect the packets to be encrypted. Furthermore, the stream will experience normal network conditions, such as packet loss, latency and jitter. We will not consider app-based calls (WhatsApp, FaceTime, etc.) because these apps can only call other users of the same app, and cannot be used to call a customer service number on the traditional phone network [11][12].

#### C. Testbed Overview

In order to make our test as realistic as possible, we use a popular VoIP equipment hardware and route the calls through the Internet. Therefore, the packet streams will experience real-world Internet latency and jitters, which will be reflected in the collected data (Fig 2). For the hardware, we use Cisco 2911 [13] with CallManager Express [14] to act as SIP server, the traditional phone line (PSTN) gateway and an IPSec tunnel [15] endpoint. As we want to expose our calls to real-world network conditions, this encrypted tunnel connection is routed from United States through Indonesia, for a total of 37 hops and an average round-trip time of 568 ms. We intentionally choose a long routing path to show that the attack is feasible even if the VoIP data stream experiences a big latency, jitter and even data loss in the path. Finally, the IPSec tunnel is terminated at a Cisco 881 Router [16] which is also connected to the Cisco 7965 [17] IP Phone (Fig 3).

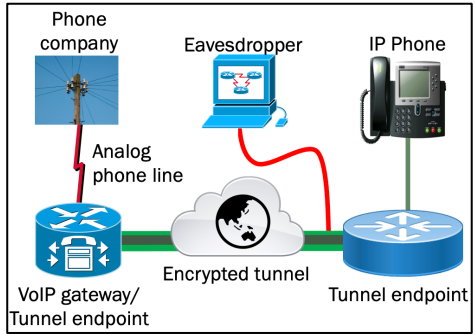

Fig. 2. Testbed topology

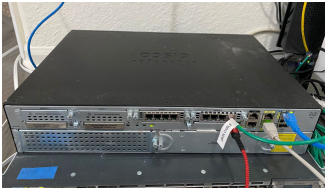

(a) Cisco 2911 as VoIP server/gateway and a tunnel endpoint

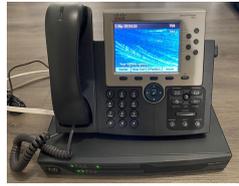

(b) Cisco 7965 as VoIP phone and Cisco 881 Router as tunnel endpoint

Fig. 3. Testbed equipment

For VoIP protocol configuration, we use the standard SIP signaling [10] (as opposed to Cisco-proprietary SCCP protocol [18]), and the default voice codec of G.729. It has 20 Bytes of voice payload per packet at 50 packets per second (i.e., 20 ms gap). For sniffing tools, we use Wireshark [19] and configure it to capture 174-Bytes packets, which is the voice payload plus the various network overhead (RTP, IPSec tunnel, IPv4, Ethernet). Even if there might be other packets that share the same packet size, we can differentiate between them because our target packets will have a consistent gap of 20 ms. A long gap in an otherwise constant 20 ms stream is indicative of the silence that we will exploit for our profiling of the call (Fig 4).

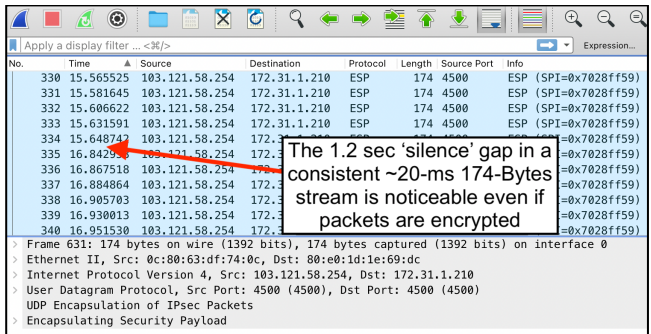

Fig. 4. Wireshark capture showing a spike in inter packet gap due to silence suppression

## IV. PROFILING AND DATA COLLECTION

For our profiling, we selected 12 popular customer service phone numbers from various industries: airlines, financial institutions, retailers, and government agencies. Wireshark captures the encrypted call stream, which will be exported and analyzed with a php script before being put in a database.

### A. VoIP Phone Dialing

Initially, we want to use programmable VoIP phone to automate the task of dialing customer service numbers. Unfortunately, we cannot find one that is compatible with our Cisco infrastructure – which we need to encrypt and route the packets. Hence, each number has to be manually dialed for every single data collection. This limits the amount of data that we can collect, fortunately we found that the speech fragment durations are very consistent, even amid high amount of network jitter due to the long network route (Fig 5).

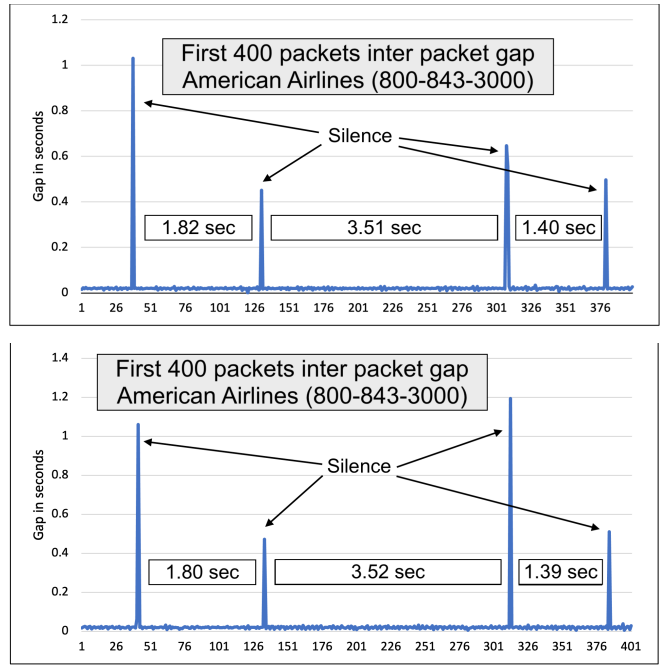

Fig. 5. Two profiling calls to American Airlines phone system, showing silence gap spikes that reveal a very consistent duration of speech fragment in between the silence

### B. Wireshark Packet Capture

We want to capture the packets associated with the call, but because we are capturing on the tunnel side, we won't be able to look inside the packet. So the filter that we use on the Wireshark is just to capture UDP packets of size 174 that is transmitted to the tunnel endpoints: <udp and ip dst "tunnel_ip" and length==174>.

The captured packets are then exported to a csv (Comma Separated Values) file so it can be analyzed by our php script.

### C. Creating Profile Database

In order to separate the speech fragments, we must first choose the threshold of gap that the script would consider as "silence". In an ideal network, any gap greater than 20 ms is a silence because the codec has a 20-ms interval. However, our attack is performed against packet streams during transit, so it must take into account the possible adverse network conditions, such as latency, jitter and packet loss. Hence, a threshold that is too small could incorrectly identify jitters and packet losses as "silence", whereas if it is too large then it might miss a real silence. In the end, we choose ten times the normal codec interval (10 x 20 ms = 200 ms) as the optimal value.

Once the fragments are identified, we need to measure their durations. We initially thought of using the number of packets, but to make the measurement more robust, we instead use the time interval between the first packet and the last packet in a fragment. By doing this, the profile will be able to tolerate packet losses in the middle of a fragment.

In our script, we also discard the first two small fragments because based on our experiments it could take up to two rings before the automated system picks up the call.

### D. Phone Number Profile Database

We run our profiler ten times for each phone number to collect the VoIP traffic. Table II summarizes the data collection result. For the first and the second voice fragments for each phone number, the table shows the minimum, maximum, and median values based on the 10 observed duration time values.

TABLE II. PROFILE OF THE FIRST TWO VOICE FRAGMENTS IN SECONDS (SORTED BY FIRST FRAGMENT DURATION)

| Phone | Company | Fragment 1 (min/max/median) | Fragment 2 (min/max/median) |
|---|---|---|---|
| 800-221-1212 | Delta Air | 1.50/1.56/1.52 | 4.33/4.35/4.34 |
| 877-383-4802 | Capital One | 2.46/2.59/2.48 | 1.99/2.14/2.14 |
| 800-841-3000 | Geico Ins | 2.64/2.78/2.66 | 2.38/2.49/2.40 |
| 800-772-1213 | Social Security | 3.22/3.24/3.23 | 3.61/3.63/3.61 |
| 800-843-3000 | American Air | 3.50/3.56/3.52 | 1.32/1.41/1.32 |
| 800-242-7338 | Chase Bank | 5.01/5.07/5.03 | 4.98/5.00/4.99 |
| 800-435-9792 | Southwest | 5.25/5.33/5.27 | 2.54/2.54/2.54 |
| 855-284-9166 | AA Credit | 5.32/5.38/5.34 | 0.99/1.84/1.17 |
| 800-375-5283 | US Immigration | 5.46/5.66/5.46 | 2.41/2.45/2.44 |
| 800-925-6278 | Walmart | 5.99/6.08/6.01 | 1.84/1.98/1.98 |
| 888-287-4637 | Bank of America | 7.45/9.31/8.52 | 4.02/4.10/4.04 |
| 866-948-8472 | United Ins | 13.82/14.74/14.15 | 2.66/2.82/2.66 |

We only collected the first two speech fragments profile because it is already enough to properly distinguish each of the phone number. As a matter of fact, most of the numbers can be identified just by looking at the first fragment. Of course there might be overlap in fragment durations if more numbers are added, and in that case additional fragments might be needed.

The result is visualized in Figure 6. The height of each bar represents the range of durations that were collected for that voice segment. Since many of the fragments have a very consistent duration, their bars on the figure are correspondingly very thin. There are three exceptions, which is for segment two of AA Credit, and segment one of Bank of America and United Ins. The case of AA Credit is unique, because after the first 5.3s fragment, the call is transferred to another system which starts with several rings before it is answered. Therefore, the variability is due to the rings, and not the speech itself. In case of Bank of America and United Insurance, we believe the first relatively long speech fragment for some reason causes this variability. But it is still distinct enough to be properly identified with the right phone number.

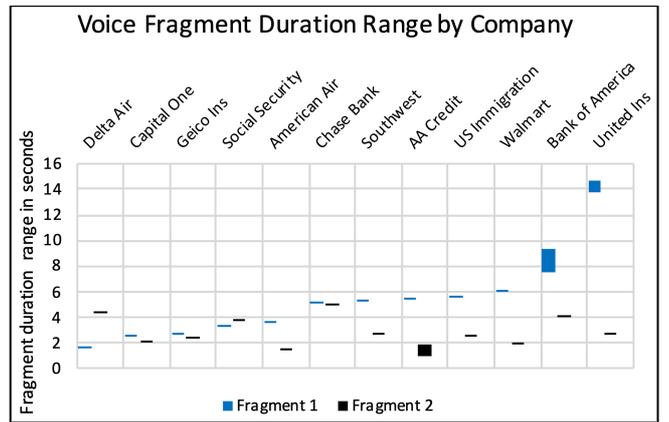

Fig. 6. Visualization of the range for the first and the second voice fragment from the 12 automated phone systems that were profiled. Virtually all fragments have a very consistent durations across the ten calls, hence the thin bars.

Some profile, such as Capital One and Geico Ins, seems to overlap on Figure 6. But if we zoom in on them, we can see that there is enough separation between them for accurate classification (Fig 7). On some profiles where the first fragment do have some overlap, such as the Southwest and AA Credit case, we can still identify them apart by looking at the second fragment, which shows clear difference (Fig 8).

The collected profile also shows that the second fragment actually has stronger one-to-one correlation to a phone number than the first fragment. However we must always start the identification process from the first because a caller might select the option (i.e., push a button) before the entire message is played by the automated phone system. Therefore, we might need to identify the call by just using the first fragment.

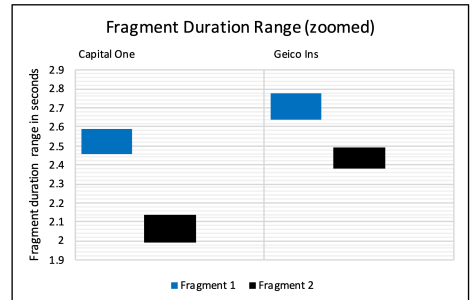

Fig. 7. Zoomed in visualization showing clear delineation between profiles that look similar in Fig 6

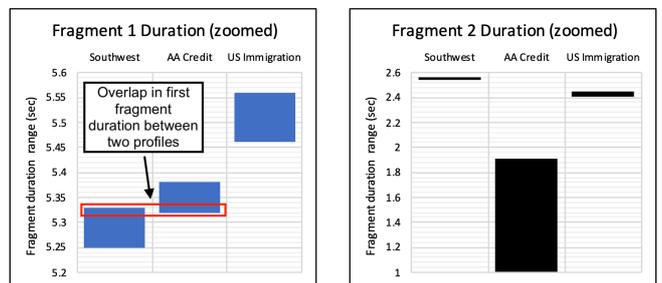

Fig. 8. An overlap in first fragment is resolved by evaluating the second fragment

Based on this, we create a simple profile database and classifier program. The database will contain a table with the phone number, its first fragment duration range and its second fragment duration range. Because each phone number has a unique profile that does not completely overlap another number, we can use a simplistic classifier and does not need to use any machine learning methods.

### E. Option Selection Profiling

Using the same data gathering method, we can also profile the options that a caller chooses during the call. This option selection profiling attack will reveal more private information of the caller, thus it is a more serious attack.

We perform this profiling on two companies. In the case of WalMart, it has multiple levels of options (Fig 9). An option is selected using touch-tone (DTMF) [20], except for the second level option under "Orders", which uses voice recognition. For demonstration, here we only show the profiling attack on the first level option selection (Table III). Whereas in Geico phone system (Table IV), it entirely uses voice recognition. For this reason, the Geico phone system sometime asks for clarification if it is not sure of the options chosen by the caller (i.e., the "home?" and "claim?" row). These clarification speech messages are difficult to reproduce, as we only experience them several times and we are unable to reproduce them consistently. Also, some Geico options have only one fragment response, hence there is no second fragment to measure.

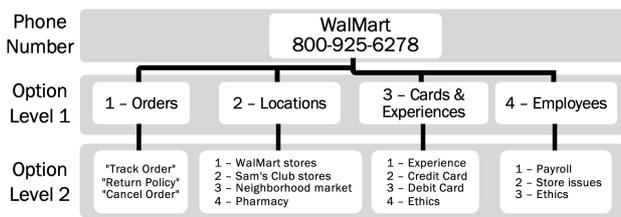

Fig. 9. WalMart Customer Service Phone Options Tree

TABLE III. WALMART TOUCH TONE OPTIONS
VOICE FRAGMENT PROFILE DATA (IN SECONDS)

| Option | Fragment 1 (min/max/median) | Fragment 2 (min/max/median) |
| --- | --- | --- |
| 1 - Orders | 2.52/2.56/2.54 | 2.20/2.27/2.24 |
| 2 - Locations | 7.63/7.98/7.65 | 2.61/2.70/2.66 |
| 3 - Cards | 2.93/3.10/2.95 | 2.55/2.78/2.58 |
| 4 - Employees | 2.81/2.86/2.84 | 3.59/3.62/3.60 |

TABLE IV. GEICO SPEECH OPTIONS
VOICE FRAGMENT PROFILE DATA (IN SECONDS)

| Option | Fragment 1 (min/max/median) | Fragment 2 (min/max/median) |
| --- | --- | --- |
| "automobile" | 1.49/1.51/1.50 | 2.28/2.30/2.30 |
| "homeowner" | 3.29/3.36/3.30 | |
| "claim" | 2.60/2.60/2.60 | 1.21/1.24/1.21 |
| home? | 2.36/2.36/2.36 | |
| claim? | 1.66/1.66/1.66 | |

## V. EVALUATION

### A. Phone Number Classification

Because there is no speech profile that completely overlaps between different companies, we can make our detection algorithm more robust. First, we increase the range of durations for any fragment profile to at least 5 times the codec interval (5 x 20 ms = 100 ms) from its median. This means we are able to tolerate a packet loss of up to five packets and a jitter of up to 5 times the normal codec interval. If the measured range is already higher than that, then we keep the higher value. For example, the second fragment duration of Delta Air (which has a median of 4.34) will become 4.24 – 4.44. Second, we can use simple if/then matching between fragment durations of the eavesdropped stream and each profile in our database. We don't need to use machine learning and its associated complexity. On the other hand, in a real attack where the database has hundreds or even thousands of profiles of customer service phone numbers, the attacker might need to rely on machine learning algorithms to do classification accurately.

As for the case of AA Credit, where the second fragment is a ringing tone, we manually change the profile for this fragment to 0.2-2s. We do this because the ringing tone could be a short one, and two-second tone is the standard ringing tone in United States [21].

Our classifier is coded in php and it will go through each phone profile to check if the current eavesdropped stream has a pattern that fits that phone's profile. If it cannot find any match after all profiles are compared, then the eavesdropped stream will be considered an "Unclassified" profile.

We run the test for five minutes for each company, which yields about 9-15 calls for each customer service phone number. The classification result is shown in Figure 10. There are two instances of incorrect identification with Capital One and AA Credit. The Capital One is due to a jitter of more than 100 ms. Whereas the AA Credit case is due to a ringing tone that goes beyond the standard two-second tone. Using the standard measurement for precision and recall [22], Capital One has 100% precision and 87% recall, whereas AA Credit has 100% precision and 92% recall. The rest of the phones all have 100% precision and recall.

Fig. 10. Confusion matrix of phone number classification

### B. Option Selection Classification

For each customer service phone number, because there are only several options (mostly no more than 9) at any level for a caller to choose, we can optimize our classifier further by only evaluating the first fragment in most cases. There is a small overlap between option 3 and 4 for Walmart, which is the only case where we need to look at the second fragment for classification. We are able to correctly identify each option chosen by the caller in the eavesdropped stream, as shown in

Figure 11 and Figure 12 for Walmart and Geico phone numbers, respectively. In both cases, the precision and recall for all of them are 100%.

|  | | Actual | | | |
|---|---|---|---|---|---|
| **Prediction** | | 1 | 2 | 3 | 4 |
| | 1 | 100 | | | |
| | 2 | | 100 | | |
| | 3 | | | 100 | |
| | 4 | | | | 100 |
| | Unclassified | | | | |

Fig. 11. Confusion matrix of WalMart touch-tone option classification for the first-level option selection

|  | Actual | | |
|---|---|---|---|
| **Prediction** | automobile | homeowner | claim |
| automobile | 100 | | |
| homeowner | | 100 | |
| claim | | | 100 |
| Unclassified | | | |

Fig. 12. Confusion matrix of Geico speech option classification

*C. Limitation*

For the attack to work, the played messages from a phone number must not change between the profiling and the eavesdropping. Any changes will require a re-profiling of the phone number in question. The attack will also fail if the caller selects the options (i.e., push the touch-tone button) prior to the system finishing the first two speech fragments. In this case, the match must be performed on partial fragment, leading to inaccuracies. Finally, if the Voice Activity Detection (VAD) is disabled, there would be no silence gap and the attack become infeasible. However, this will also eliminate the bandwidth saving benefit associated with VAD.

## VI. CONCLUSION

Our paper demonstrated that by exploiting the static speech pattern of automated customer service phone, an attacker can potentially reveal private information about a caller, including which customer service phone number (and thus the company) the call is to, and even the subsequent options chosen by the caller. Furthermore, once the correct profile is obtained, the attack is very accurate with many instances of 100% precision and recall.

For future work, we can study the same vulnerability in IVR where the personal information provided by a caller is repeated back by the automated phone system, in most cases, to let the caller confirming the accuracy of he or she inputs. If the same profiling can be performed on such IVR response, this could lead to a much more serious privacy leakage of the caller's personal information such as birth date, phone number, or account number.